\documentclass[twocolumn,5p,times]{elsarticle}

\usepackage{amssymb}

\usepackage{soul}
\usepackage{color}

\begin{document}

\begin{frontmatter}

\title{Energy reconstruction of hadron-initiated showers of ultra-high energy cosmic rays}

\author[ALCALA]{G. Ros}
\ead{german.ros@uah.es}
\author[UNAM]{G. A. Medina-Tanco}
\ead{gmtanco@nucleares.unam.mx}
\author[IAFE]{A. D. Supanitsky}
\ead{supanitsky@iafe.uba.ar}
\author[ALCALA,ITEDA]{L. del Peral}
\author[ALCALA,ITEDA]{M. D. Rodr\'iguez-Fr\'ias}
\address[ALCALA]{Space and Astroparticle Group, Dpto. F\'isica y Matem\'aticas, Universidad de Alcal\'a.
Ctra. Madrid-Barcelona km. 33. Alcal\'a de Henares, E-28871, Spain.}
\address[UNAM]{Instituto de Ciencias Nucleares, UNAM, Circuito Exteriror S/N, Ciudad Universitaria,
Mexico D. F. 04510, Mexico.}
\address[IAFE]{Instituto de Astronom\'ia y F\'isica del Espacio, IAFE, CONICET-UBA, Argentina.}
\address[ITEDA]{ITeDA (CNEA CONICET - UNSAM) - Buenos Aires, Argentina.}
\begin{abstract}

The current methods to determine the primary energy of ultra-high energy cosmic rays (UHECRs) are different when dealing
with hadron or photon primaries. The current experiments combine two different techniques, an array of surface detectors 
and fluorescence telescopes. The latter allow an almost calorimetric measurement of the primary energy. Thus, hadron-initiated 
showers detected by both type of detectors are used to calibrate the energy estimator from the surface array (usually the 
interpolated signal at a certain distance from the shower core $S(r_0)$) with the primary energy. On the other hand, 
this calibration is not feasible when searching for photon primaries since no high energy photon has been unambiguously 
detected so far. Therefore, pure Monte Carlo parametrizations are used instead.

In this work, we present a new method to determine the primary energy of hadron-induced showers in a hybrid experiment based on a technique 
previously developed for photon primaries. It consists on a set of calibration curves that relate the surface energy estimator, $S(r_0)$, and the
depth of maximum development of the shower, $X_{max}$, obtained from the fluorescence telescopes. Then, the primary energy can be determined from 
pure surface information since $S(r_0)$ and the zenith angle of the incoming shower are only needed. Considering a mixed sample 
of ultra-high energy proton and iron primaries and taking into account the reconstruction uncertainties and shower to shower fluctuations, 
we demonstrate that the primary energy may be determined with a systematic uncertainty below 1$\%$ and resolution around 16$\%$ in the energy range 
from 10$^{18.5}$ to 10$^{19.6}$ eV. Several array geometries, the shape of the energy error distributions and the uncertainties due to the unknown 
composition of the primary flux have been analyzed as well.
\end{abstract}

 
\end{frontmatter}

\section{Introduction} 

\label{Sec:Intro}

The energy spectrum of cosmic rays extends by more than 10 orders of magnitude from below 1 GeV to more than 10$^{20}$ eV. The energy spectrum
follows a power law as $E^{-\gamma}$, where $\gamma$ is around 3.0 in the whole energy range. It is so steep that direct measurements are not 
feasible above 100 TeV. At higher energies, the properties of the primary cosmic ray are determined indirectly from the measurement of the 
extensive air shower (EAS) it produces after colliding with molecules of the atmosphere.

The highest energy EASs have been traditionally studied using two different techniques. The first one is based on telescopes that collect the 
fluorescence light emitted by atmospheric Nitrogen molecules excited by secondary particles of the EAS (e.g., Fly's Eye, HiRes). This allows to determine 
the longitudinal profile of the shower and it is considered to be close to a calorimetric measurement of the UHECR primary energy. However, fluorescence
light can only be observed during moonless nights and, consequently, this technique can only be applied to $\sim13\%$ of the incoming events 
\cite{AugerFD:10}. The second technique involves an array of detectors located at ground level, mainly scintillators (e.g., Volcano Ranch, AGASA, KASCADE)
or water Cherenkov tanks (e.g., Haverah Park), whose duty cycle is close to $100\%$. Thus, the lateral distribution of secondary particles at ground 
level can be inferred from the discrete sampling of the shower front. The lateral distribution is fitted assuming an appropriate parametrization
(called the lateral distribution function, LDF). The interpolated signal at a certain optimum distance, $S(r_0)$, is used as the energy estimator, which 
can be related to the primary energy thorough, for instance, Monte Carlo (MC) parametrizations. The optimum distance, $r_0$, is traditionally fixed for 
each detector since it is assumed to be only dependent on the array spacing and geometry \cite{Newton:07}, although some studies suggest the 
convenience  of calculating the optimum distance for each individual shower taking into account its primary energy and direction \cite{Ros:09}.

The current experiments, on the other hand, use hybrid techniques for $S(r_0)$ calibration. The Pierre Auger Observatory \cite{AugerNIM:04}, 
taking data since 2004, pioneers the simultaneous use of water Cherenkov detectors and fluorescence telescopes, while the Telescope Array Observatory
\cite{TADescription:07}, operating since 2008, combines scintillators and telescopes. Events detected simultaneously by both the surface and 
fluorescence detectors are called {\it hybrid}. Hybrid events allow the calibration of $S(r_0)$ with primary cosmic ray energy \cite{AugerICRCSpectrum:13}.
Thus, the energy of each event detected by the surface detector alone can be determined almost independently of MC simulations. Systematic errors in
energy estimate are greatly reduced in this way \cite{AugerSprectrum:10, TASDSpectrum:13}. These calibrations assume that the primaries are nuclei and, 
therefore, they cannot be directly applied to photon-initiated showers. In addition, no photon event has been unambiguously identified up to now by 
any experiment so a proper calibration for photons is not possible with this technique. Therefore, each experiment relies on MC simulations to infer 
the primary energy of photon events \cite{HPPhoton:02, AGASAPhoton:02, AGASAPhoton:05, TAPhoton:13, AugerPhoton:08}. 

The method used for photon searches by Auger in Ref. \cite{AugerPhoton:08} was first proposed in Ref. \cite{Billoir:07}. This method takes into account 
the well-known universality of the electromagnetic component of EAS \cite{Lafebre:09,Kascade:08,Schmidt:07} and the small muon fraction of the photon-initiated 
showers. The calibration curve, that is obtained from MC simulations, relates $S(r_0)$, the zenith angle of the incoming shower, $\theta$, and $X_{max}$. Thus, 
the primary energy of photon primaries can be determined with resolution of $\sim$20-25$\%$ \cite{AugerPhoton:08,Billoir:07}. 

In this work, we show how to modify that method to be applicable to hadron-initiated showers where the muon component is significant, especially, 
in case of water Cherenkov arrays which enhanced their contribution to the total measured signal. The additional advantage is that the same method 
could be used to infer the primary energy for both, photon and hadron showers. Moreover, in case of hadron-initiated showers the method can
be calibrated with hybrid events reducing the systematic uncertainties coming from the high energy hadronic models used for shower simulations. 

\section{Shower and detector simulations} 
\label{Sec:Simulations}
The simulation of the atmospheric showers is performed with the AIRES Monte Carlo program (version 2.8.4a) \cite{Aires:99} using QGSJET-II-03 
\cite{QGII:06} as the hadronic interaction model. The input primary energy goes from $\log(E/eV)$ $=$ $18.5$ to $19.6$ in $0.1$ steps. 
Approximately $2000$ events have been simulated per energy bin for both, proton and iron primaries. The zenith angle has been selected following 
a sine-cosine distribution from $0$ to $60$ degrees, while the azimuth angle is uniformly distributed from $0$ to $360$ degrees. $X_{max}$ is 
obtained from these simulations.

Given the energy, the zenith and azimuth angles of the shower, the detector response is simulated with our own code, previously tested in
Refs. \cite{Ros:09,Ros:11,Ros:13}. Following the original proposal in Ref.~\cite{Billoir:07}, we select a triangular array of cherenkov detectors 
separated 1.5 km and $S(r_0=1000 \textrm{m}) \equiv S(1000)$ as the energy estimator. The real core is randomly located inside an elementary cell
while the reconstructed core position is determined by fluctuating the real one with a Gaussian function whose standard deviation depends on the 
primary energy, composition and the distance between detectors (see Ref.~\cite{Ros:09} for more details). 

The signal collected at each station for a given shower is set assuming a \textit{true} lateral distribution function of the form,
\begin{equation} \label{AugerLDF}
S(r)=S(1000)\times\left(\frac{r}{r_0}\right) ^{-\beta}\times\left( \frac{r+r_s}{r_0+r_s}\right)^{-\beta},
\end{equation}
where $r_s$ = 700 m, $r_0$=1000 m, the distance to the shower axis $r$ is in meters, $S(1000)$ is in VEM (vertical equivalent muons,
unit for the energy deposited by a vertical muon in a water tank \cite{AugerNIM:04}) and $\beta(\theta,S(1000))$ is given by 
(based on work by T. Schmidt et al. \cite{Schmidt_GAP:07} as presented in \cite{Maris_Thesis}),
\begin{equation} \label{BetaAugerLDF}
\beta(\theta,S(1000)) = \left\lbrace
  \begin{array}{ll}
   a + b (\sec\theta - 1) & \textrm{if} \sec\theta < 1.55 \\
   a + b (\sec\theta - 1) & \\
   + f (\sec\theta -1.55)^2  & \textrm{if} \sec\theta > 1.55   
  \end{array}
  \right. 
\end{equation}
where $a = 2.26+0.195 \log(e)$, $b = -0.98$, $c = 0.37-0.51 \sec\theta+0.30 \sec^2\theta$, $d = 1.27-0.27 \sec\theta + 0.08 \sec^2\theta$, 
$e=c\ S(1000)^d$ and $f = -0.29$. 

A realistic $S(1000)$ to be used in Eqs.~(\ref{AugerLDF}) and (\ref{BetaAugerLDF}) is obtained from,
\begin{eqnarray}\label{CICHybrid}
E &=& A\ (S_{38})^B,  \nonumber \\
S(1000)(\theta)&=&S_{38}\times\left[1+ Cx - Dx^2 \right],  
\end{eqnarray}
where $x = \cos^2(\theta)-\cos^2(38^o)$. $A$, $B$, $C$ and $D$ are constants given in Ref. \cite{Maris_Thesis} for QGSJetII-03, iron and 
proton primaries. In addition, shower to shower fluctuations for each primary are emulated by fluctuating the value from Eq.~(\ref{CICHybrid}) 
with a Gaussian distribution whose standard deviation is taken from Fig.~3 in Ref.~\cite{AveAugerICRC:07}. 

Finally, the signal assigned to each station is fluctuated using a Poissonian distribution whose mean is given by the \textit{true} LDF. 
We adopt $S_{th} = 3.0$ VEM and $S_{sat} = 1221$ VEM as trigger and saturation thresholds respectively \cite{Ros:09}. 

Next, the lateral distribution of particles is fitted using a functional form given by,
\begin{equation}\label{AugerLDFFit}
\log S(r)=a_1+a_2\left[\log\left(\frac{r}{r_0}\right)+\log\left(\frac{r+r_s}{r_0+r_s}\right)\right],
\end{equation}
where the slope of the LDF and the normalization constant are free parameters while the core position is fixed in the reconstructed one. The values
of $\chi^2$/\textit{ndf} are good if at least 3 stations are included in the fit, a minimum condition for shower reconstruction that is fulfilled for
almost every event above the energy threshold of the detector. Finally, the reconstructed $S(1000)$ is determined as the interpolated value from 
the fit at $1000$ meters from the shower axis. In this method, event by event fluctuations and reconstruction uncertainties are properly taken 
into account.

The problem of saturation is common to all surface arrays, specially when dealing with high energy vertical showers. The consequent lack of detectors 
close to the core produces large uncertainties in the LDF fit and affects the reconstructed $S(r_0)$. The Auger Collaboration, for example, has 
developed sophisticated algorithms to estimate the signal of a saturated detector \cite{AugerICRCSaturation:13}. Nevertheless, the analysis of such 
uncertainties and how to minimize them is beyond the scope of the present work so saturated events are discarded here.

The simulation set has been divided into two samples. In each sample, an equal number of proton and iron primaries have been mixed for each 
energy bin. The first sample represents the hybrid events and it is used to determine the calibration curves as it will be explained in the next Sections. 
Typical values for their reconstruction uncertainties are considered, so their real energy, zenith angle and $X_{max}$ are fluctuated with Gaussian 
distributions whose standard deviations are 15$\%$ \cite{AugerICRCSpectrum:13,TASDSpectrum:13}, 1$^o$ \cite{AugerAngular:09, TAHybridSpectrum:13} and 
20 g/cm$^2$ \cite{CompositionGroup:13} respectively. The second sample, which represents data from the surface detector alone, is used for reconstruction 
and only their reconstructed S(1000) and zenith angle are needed. Thus, to simulate the reconstructed $\theta$ we have proceeded as previously for the 
calibration sample. Note that the method proposed here is obviously also applicable to pure surface arrays but the calibration should be performed with 
MC simulations in this case as, in fact, always occurs in these type of experiments.

\section{Method} 
\label{Sec:Method}
The basic idea is that the dependence of $S(r_0)$ with energy and zenith angle can be factorized as $S(r_0) = E^\alpha f(\theta)$, where $\alpha$ 
is slightly less than 1 (for example 0.95 in Refs. \cite{AugerNIM:04,Billoir:07}). $f(\theta)$ takes into account the longitudinal evolution of the shower
so it should be a decreasing function of $\theta$ and it depends on the slant depth of the shower $X = X_{ground}/\cos(\theta)$, where $X_{ground}$ 
is the atmospheric depth at ground level. $f(\theta)$, as a function of $X$, behaves similarly to the global profile of the shower. Thus, as first 
approximation, $f(\theta)$ is a function of $X-X_{max}$ with a similar shape to the Gaisser-Hillas function commonly used to describe the 
longitudinal profile. An empirical parametrization is given in Ref.~\cite{Billoir:07} by,
\begin{equation}\label{Eq:S1000_over_E}
\frac{S(1000)}{E} = p_0 \times \frac {1 + \frac{\Delta X - 100}{p_1}} {1 + \left(\frac{\Delta X - 100 }{p_2}\right)^2},
\end{equation}
where $\Delta X = X - X_{max}$, $p_1$ and $p_2$ are in g/cm$^2$, $S(1000)$ is in VEM, the energy is in EeV and $p_0$ is in VEM/EeV. In 
the case of photon showers, whose muon contribution to the signal could be neglected and considering the universality of the electromagnetic 
component of EAS, this function results nearly independent of the primary energy. In fact, in Ref. \cite{Billoir:07} a universal parametrization 
is found with $p_0 = 1.4$ VEM/EeV, $p_1 = 1000$ g/cm$^2$ and $p_2 = 340$ g/cm$^2$. 

However, the profile does depend on the primary energy for hadron-initiated showers mainly due to the existence of a non-negligible muonic 
component in the showers, which is itself a function of the primary energy. Moreover, in case of water Cherenkov detectors the sensitivity to muons 
is enhanced. Therefore, despite the fact that the shape of the calibration curve is dominated by the photon component of the signal, the muon component 
breaks the universality. In fact, if a unique function were applied to determine the primary energy of hadron-induced showers, it would result in an 
energy dependent bias, as it will be shown later in Sec. \ref{Sec:Results:EnergyReconstruction}. Therefore, the parameters $p_0$, $p_1$ and $p_2$ 
should be allowed to change with the primary energy to correctly reproduce the profile of hadron showers. Then, $p_1$ tends to be larger than 
1000 g/cm$^2$, usually fluctuating around 3000 g/cm$^2$. In fact, the function is very slightly modified if $p_1$ is larger than 1000 g/cm$^2$ 
(the numerator is very close to unity), so we have decided to fix $p_1$ = 3000 g/cm$^2$. Then, $p_0$ is the maximum of the function and decreases 
as energy increases. Finally, $p_2$ is related to the width of the function, decreasing smoothly as energy increases.

On the other hand, $X_{max}$ in Eq.~(\ref{Eq:S1000_over_E}) can be obtained from its average dependence on energy,
\begin{equation}\label{Eq:Xmax_vs_Energy}
X_{max} = q_0 + q_1 \times \log(E/EeV).
\end{equation}
where $q_0$ and $q_1$ are in g/cm$^2$. 

Therefore, we propose in this work to obtain from the hybrid events the next calibration curves:
\begin{itemize}
 \item[]\hspace{-0.6cm}(A) the global curve using all these events following Eq.~(\ref{Eq:S1000_over_E}),
 \item[]\hspace{-0.6cm}(B) a set of curves, one for each energy bin, following Eq.~(\ref{Eq:S1000_over_E}) in order to account properly for the muonic 
component of the showers. It is important that these curves do not cross and that the statistics is good enough to  assure a good fit near 
their maximum,
 \item[]\hspace{-0.6cm}(C)  the X$_{max}$ evolution as a function of energy following Eq.(\ref{Eq:Xmax_vs_Energy}),
\end{itemize}
where the parameters needed are $S(1000)$, the zenith angle, the reconstructed energy and X$_{max}$. They could be obtained from the standard fluorescence 
reconstruction and the LDF fit. These curves, obtained from simulations, are shown in Fig.~\ref{Fig:CalibCurves} (more details in 
Sec.~\ref{Sec:Results:Calibration}).

Then, given a pure surface event, its energy could be determined from the reconstructed $S(1000)$ and the zenith angle of the incoming shower. 
Both can be obtained from the LDF fit and the geometrical reconstruction respectively. The procedure is as follows:
\begin{enumerate}[(1)]
\item Using an initial estimation of the primary energy (5 ,10 or 30 EeV), X$_{max}$ is obtained with (C). As it will be shown later, the reconstructed
energy do ºnot depend on this choice.
\item X$_{max}$ is used to get an energy estimation using (A).
\item This energy is used to get X$_{max}$ again with (C).
\item Steps (2) and (3) are repeated until the difference between two consecutive energies converges to a stable value ($\Delta E / E < 10^{-5}$). 
Around 3-5 iterations are required.
\end{enumerate}
At this step, a reconstructed energy is obtained, but as explained before, the non-universality of (A) for hadron primaries introduce a
significant bias that must be corrected.
\begin{enumerate}[(1)]
\setcounter{enumi}{4}
\item The previous reconstructed energy is used to select the nearest calibration curve from the set (B).
\item Steps (1) to (4) are repeated using the new curve from (B) instead of (A), so a new reconstructed energy is obtained.
\item If the nearest curve from (B) to the new reconstructed energy is the same as before, the process finishes. Otherwise, (5) and (6)
are repeated. Only 2 or 3 different calibration curves are usually needed.
\end{enumerate}
Following this procedure the energy bias is corrected as it will be shown later in Sec. \ref{Sec:Results:EnergyReconstruction}. 

In a real experiment, the events with a large error in the reconstructed zenith angle should be analyzed carefully or even rejected since the process 
could not converge. In fact, they are mostly saturated events with energy very close to the threshold of the detector, or events with cores very 
close to the border of the array or to a detector that is not working, so they are in general already rejected by the quality cuts usually imposed for 
data analysis.

\section{Results}
\label{Sec:Results}
\subsection{Calibration}
\label{Sec:Results:Calibration}

\begin{figure}
\centering
\includegraphics[width=8.1cm]{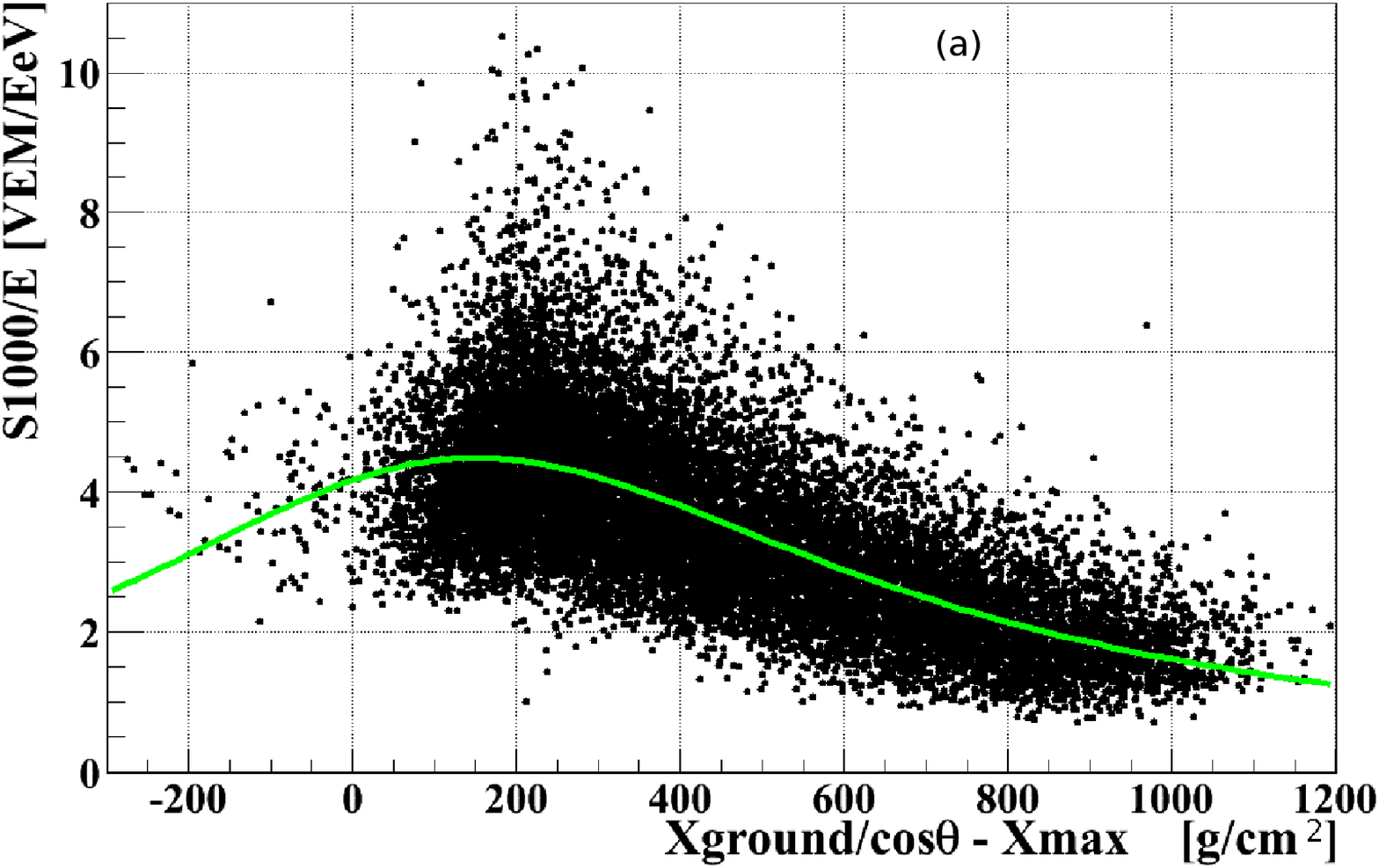}
\includegraphics[width=8.1cm]{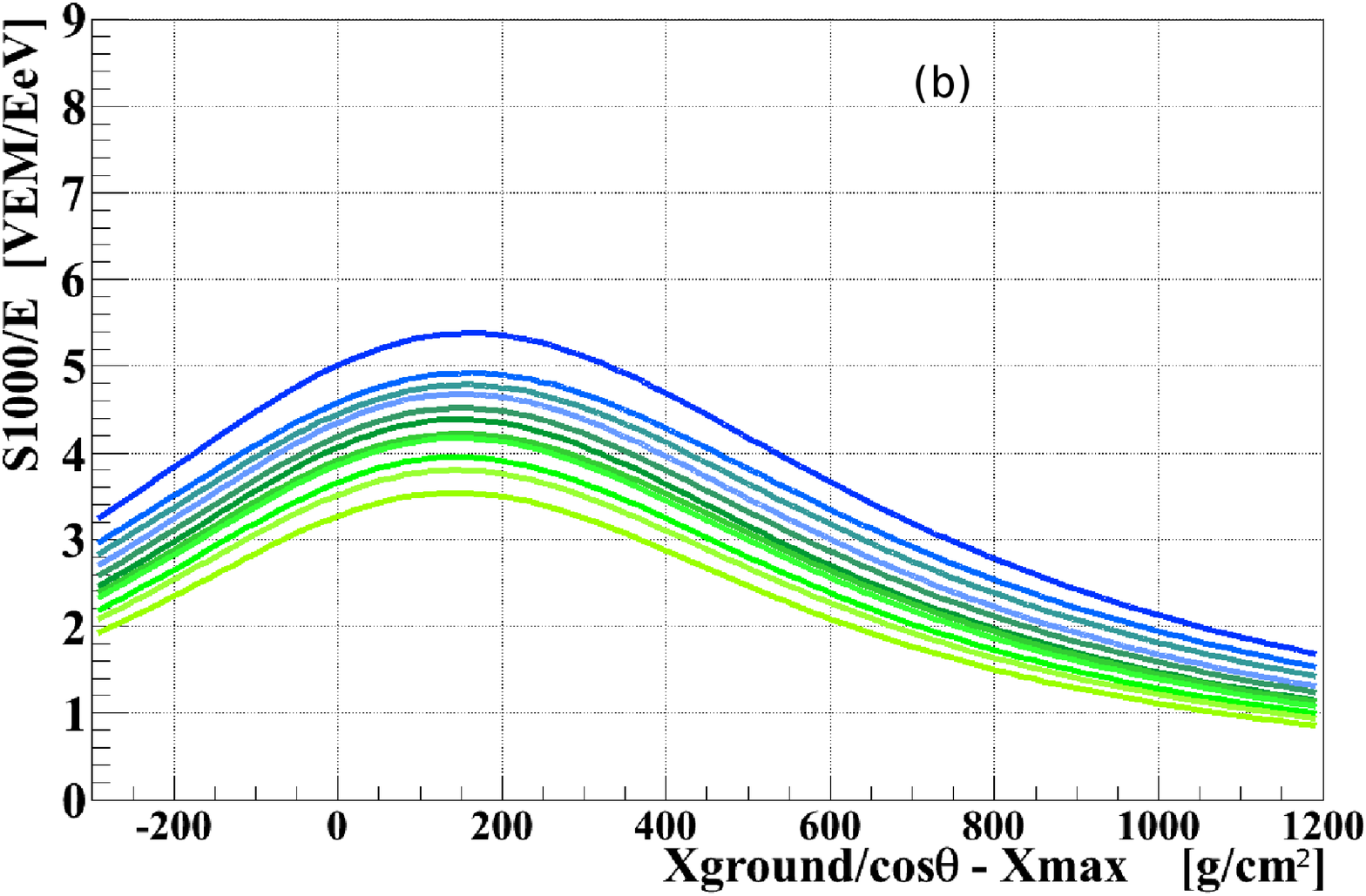}
\includegraphics[width=8.1cm]{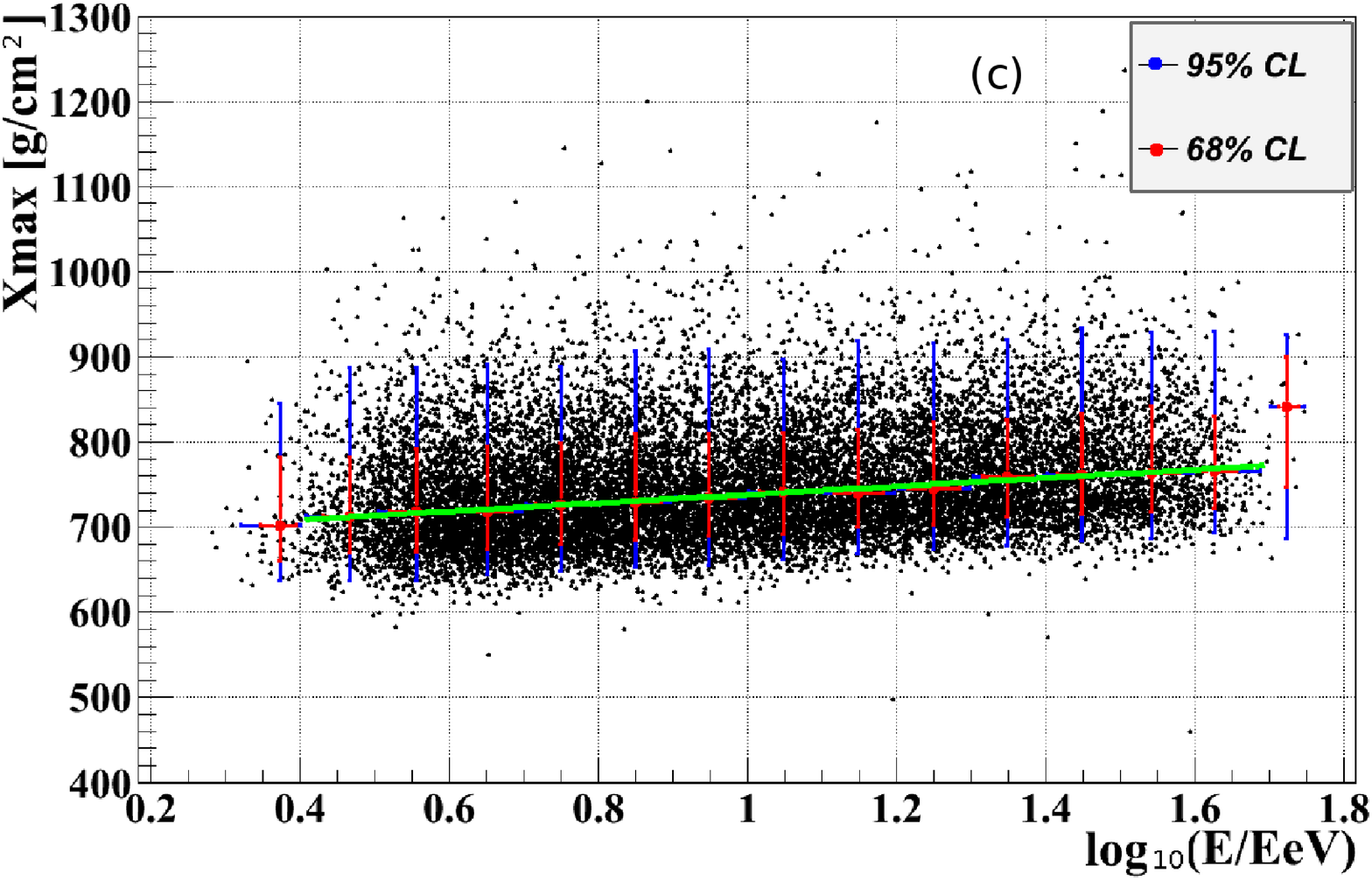}
\caption{Calibration curves for a mixed sample of iron and proton primaries. The global fit (a) and a different fit for each energy bin (b) are shown. 
In (b) the curve is lower as energy increases. The evolution of $X_{max}$ as a function of energy is shown in (c). See Sec.\ref{Sec:Method} for details.}
\label{Fig:CalibCurves}
\end{figure}

Independent sets of calibration curves can be, and indeed were, obtained for each primary. Nevertheless, we only show here the results for an equal
mixture of proton and iron primaries. The global curve and the fits for each energy bin 
($\Delta \log(E/\textrm{eV})=0.1$) are presented in Fig.~\ref{Fig:CalibCurves}.a and \ref{Fig:CalibCurves}.b respectively. The higher the energy, 
the lower the curve. Parameters $p_0$ and $p_2$ are free while $p_1$ is fixed at $p_1=3000$ g/cm$^2$ as previously explained. $p_0$ and $p_2$ smoothly 
decrease as energy increases. The evolution of $X_{max}$ as a function of energy is shown in Fig.~\ref{Fig:CalibCurves}.c, where the medians of the 
distributions are fitted taking into account the corresponding confidence levels shown in the figure. Note that the first and the last energy bins 
are not included since both are unavoidably affected by the limited energy range of the simulation set. Therefore, despite the fact that their corresponding 
curves are shown in Fig.~\ref{Fig:CalibCurves}.b, neither of them is shown in the remaining plots.

\subsection{Energy reconstruction}
\label{Sec:Results:EnergyReconstruction}

\begin{figure}
\centering
\includegraphics[width=8.1cm]{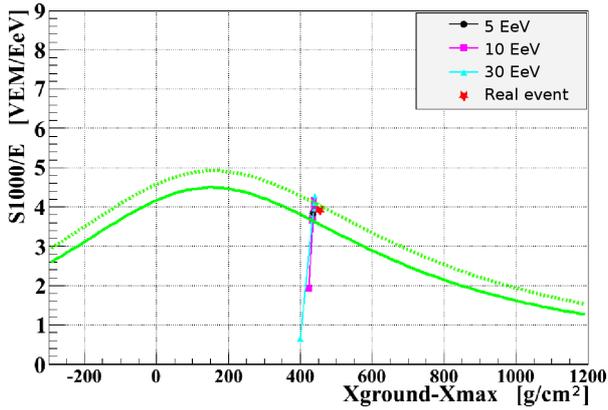}
\caption{An example of the reconstruction method: first, the global calibration curve (solid line) allows to get a first estimation of the primary energy 
by iterating between Eq.(\ref{Eq:Xmax_vs_Energy}) and itself (process not shown). This energy is used to select a new calibration curve (dashed line) 
repeating the iterative process with the new curve instead of the global curve. The path followed during the iterative process is shown in the Figure.
As can be seen, the convergence point is the same independently of the initial value used to start the iteration (5, 10, 30 EeV) and it is very close to 
the real position of the event (red star). In this example, the error in the final reconstructed energy is -5$\%$ while it would be larger than +30$\%$ 
if only the global curve were used.}
\label{Fig:IterativeProcess}
\end{figure}

We use the reconstruction sample, which is statistically independent from the calibration, in order to test the method. Given the reconstructed $S(1000)$ 
and zenith angle of each event, its energy is determined. As an example, Fig.~\ref{Fig:IterativeProcess} shows the iterative process for a typical 
event. Solid line represents the global calibration curve while the dashed line is the one used in the last step of the process. It can be seen that
the latter is much closer to the real position of the event (red star), improving significantly the energy reconstruction. It was verified that the 
convergence point is independent of the path followed, so the reconstructed energy does not depend on the initial value used to start the iteration 
as previously mentioned.

\begin{figure}
\centering
\includegraphics[width=8.1cm]{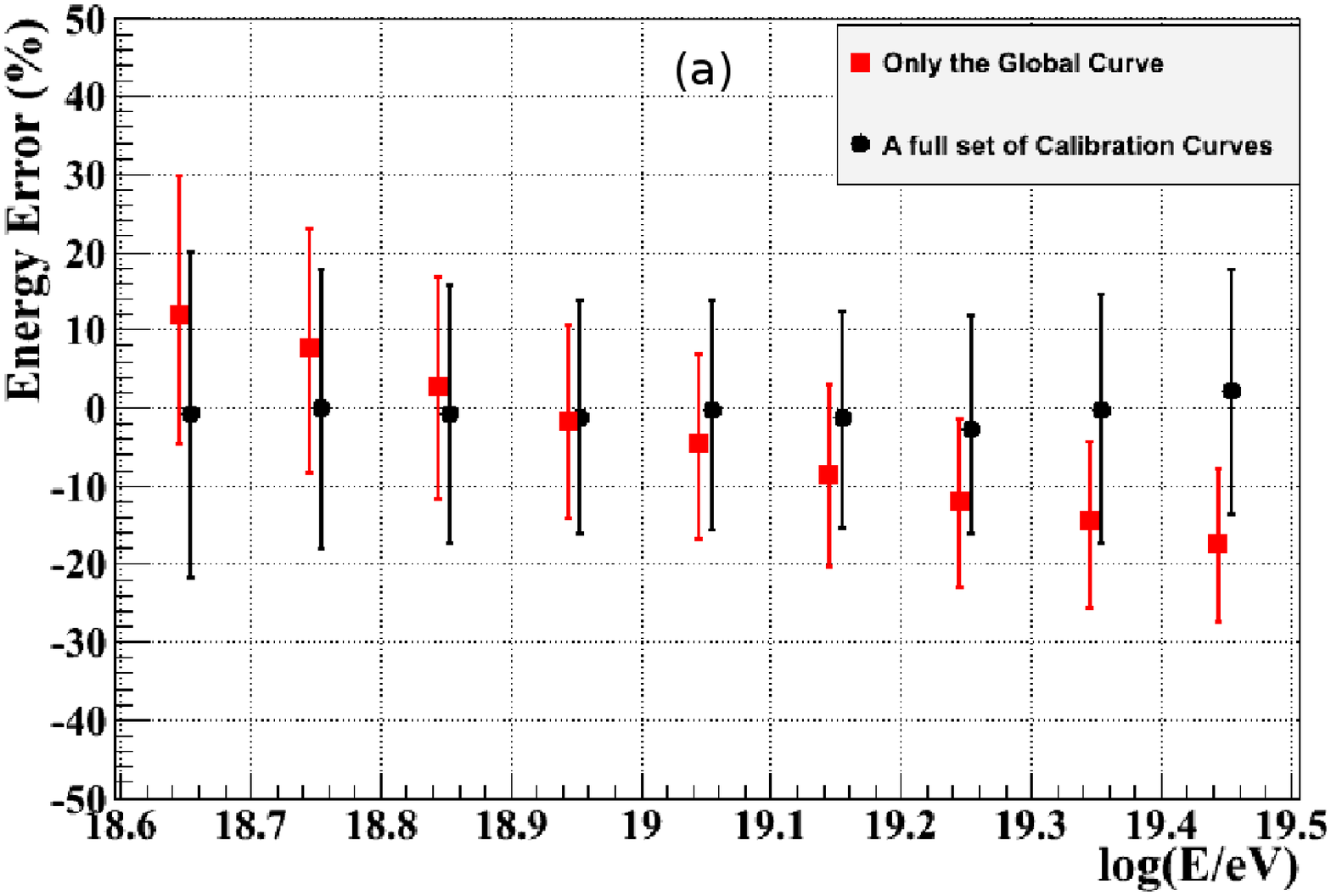}
\includegraphics[width=8.1cm]{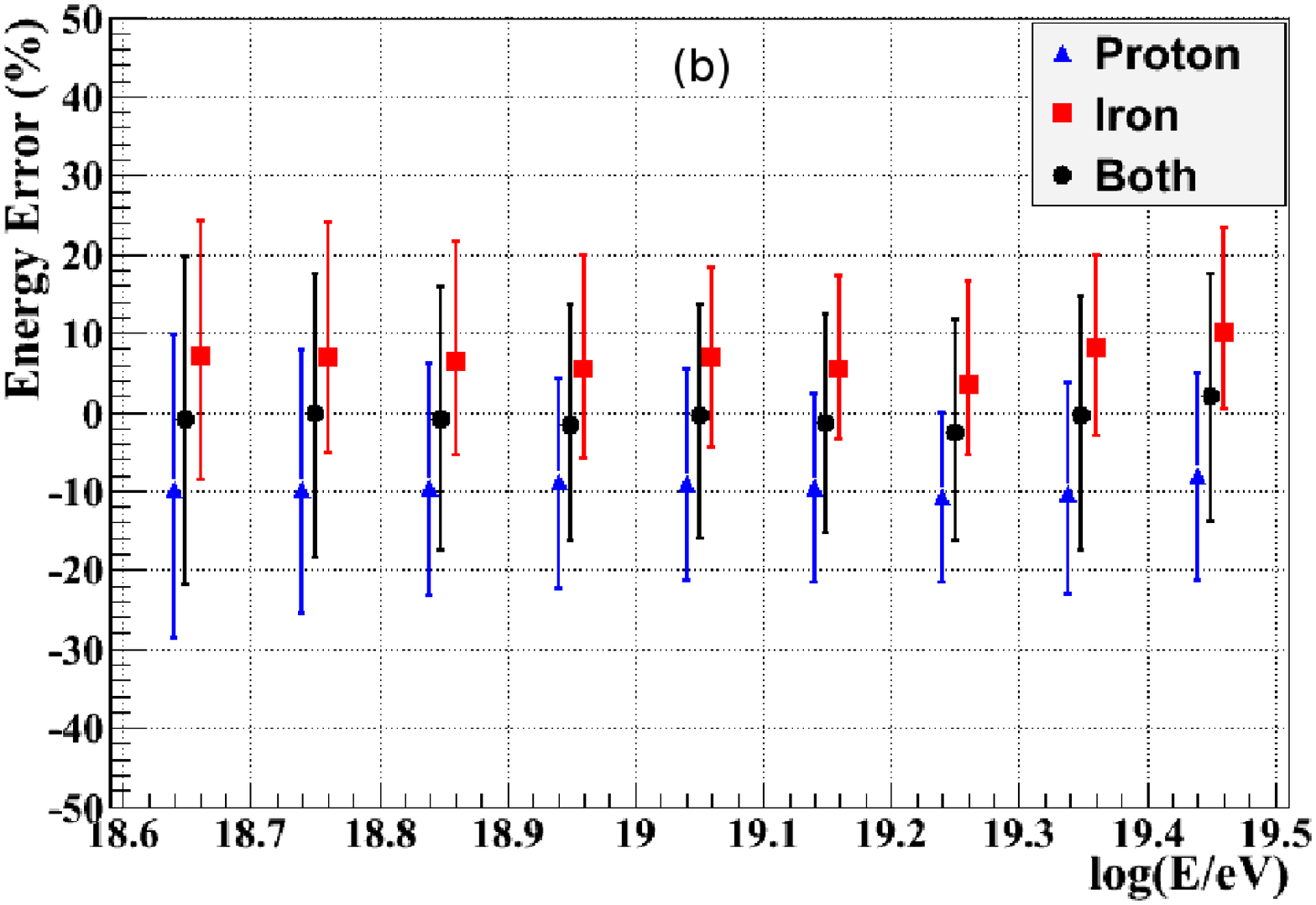}
\caption{Energy error as a function of primary energy. The points and the error bars are the median and the region of 68$\%$ probability, respectively. 
(a): Using only a global calibration (red squares) or using a full set of calibration curves (black circles) for a mixture of iron and proton primaries. 
(b): Proton (blue triangles) and iron (red squares) taken separately and together (black circles) are shown.}
\label{Fig:ErecBias}
\end{figure}

\begin{figure}
\centering
\includegraphics[width=8.1cm]{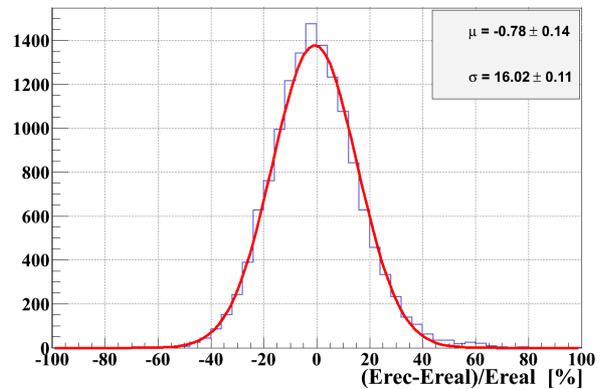}
\caption{Energy error distribution for the whole sample where equal number of iron and proton primaries in each energy bin are selected.}
\label{Fig:ErecDistribution}
\end{figure}

Fig.~\ref{Fig:ErecBias} shows the error in the reconstructed energy. The two bins at the edges have been rejected for the same reasons as before.
Fig.~\ref{Fig:ErecBias}.a shows that the energy dependent bias resultant from the use of only the global calibration curve is greatly reduced by 
employing a set of calibration curves for discrete energies, although a residual error $\leq$1$\%$ still remains. Fig.~\ref{Fig:ErecBias}.b also 
shows the error for proton and iron primaries. The error is almost energy independent for both, each primary taken separately and also for 
the mixed composition sample. Note also that the error bars, which can be considered as the resolution of the method, 
are also slightly energy dependent. The energy error distribution for both primaries taken together is approximately Gaussian and the resolution of 
the method is around 16$\%$, as shown in Fig.~\ref{Fig:ErecDistribution}. In the case of Iron and proton taken separately, the distributions are also 
nearly Gaussian and the resolution is around 13.5$\%$ and 14.5$\%$ respectively.


It is important to note that the method is quite sensitive to the $X_{max}$ vs. $\log(E/\textrm{EeV})$ calibration curve. In fact, if the parameter 
$q_0$ in Eq.~(\ref{Eq:Xmax_vs_Energy}) is modified by $\pm$10$\%$, an energy error of $\mp$7$\%$ will be produced. This effect is not so important  for 
$q_1$, since if it changes $\pm$10$\%$, a negligible error of $\mp$1$\%$ will be obtained.

\subsection{Gaussianity of the energy error distributions}
\label{Sec:Results:Guassianity}
Arguably, it is desirable that the errors in energy reconstruction follow a Gaussian distribution. Gaussian errors, for example, are easier to handle 
and understand when applying deconvolution techniques for determination of the spectrum while assuring that there are no asymmetries or long tails, 
which further reduces the danger of the limited energy range of the detector and biases associated with a rapidly changing spectral index. An example 
of these undesirable effects can be seen in Ref. \cite{Ros:09}.

\begin{figure}
\centering
\includegraphics[width=8.1cm]{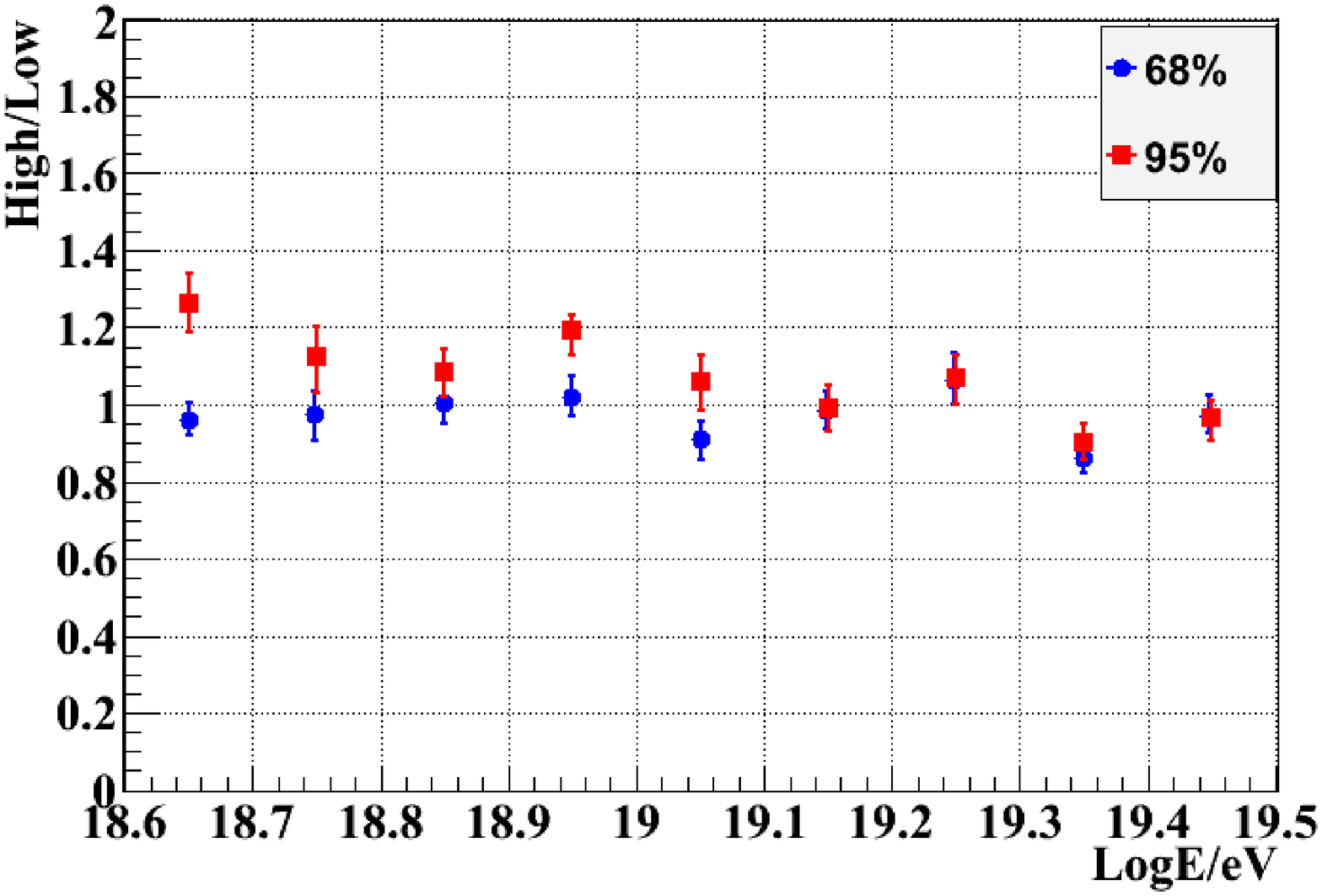}
\includegraphics[width=8.1cm]{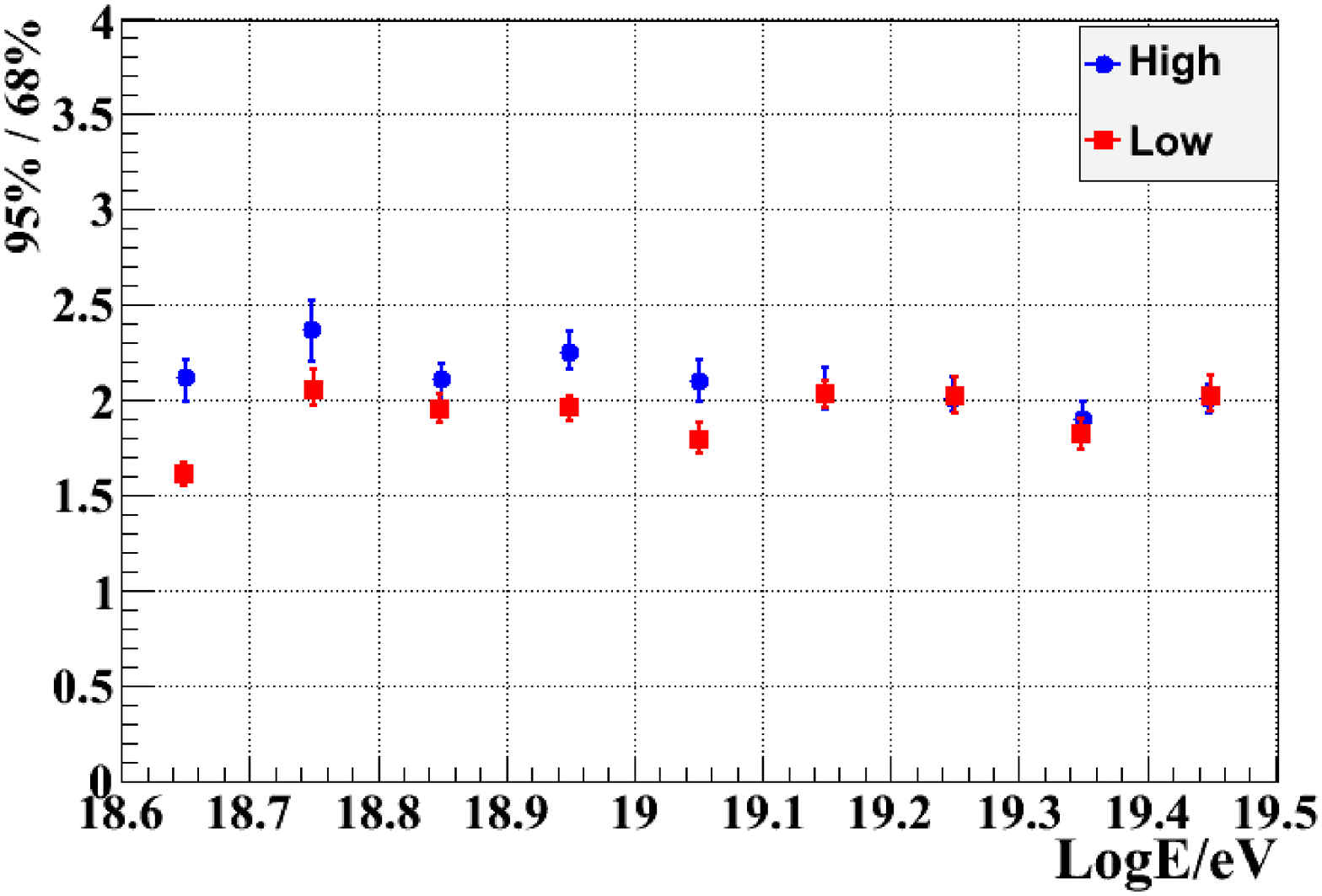}
\caption{Analysis of the symmetry (top) and the tails (bottom) of the energy error distributions. See text for details.}
\label{Fig:Gaussianity}
\end{figure}

In case of a Gaussian distribution, the ratio between the high and low parts of the $68\%$ and $95\%$ confidence levels (C.L.) should be 1 since 
the distribution is symmetric. In addition, the ratio between the $95\%$ and $68\%$ C.L., for both the high and low parts, should be 2 since they 
represent the values for 2$\sigma$ and 1$\sigma$ respectively. Fig.~\ref{Fig:Gaussianity} shows both ratios for the energy error distribution 
obtained in each energy bin. It has been calculated using the bootstrap technique that consists on resampling the original distribution a large 
number of times and calculating these ratios for each new sample. The median an $68\%$ C.L. for each ratio are shown in the plots. It can be 
seen that the error distributions are nearly Gaussian with a very good symmetry and the absence of long tails.

\subsection{Energy bias as a function of the primary composition}
\label{Sec:Results:BiasCompositon}
\begin{figure}
\centering
\includegraphics[width=8.1cm]{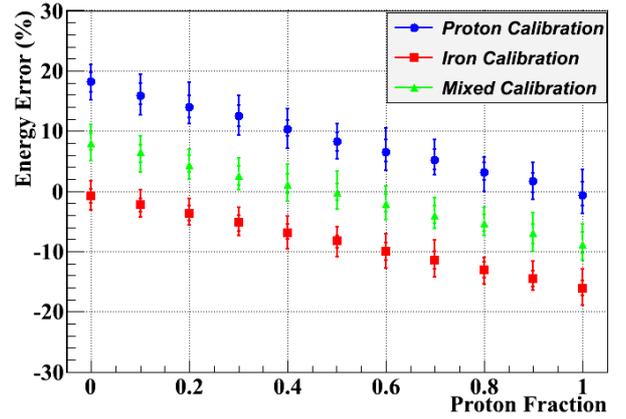}
\caption{Energy error as a function of the proton fraction of the reconstructed sample using three different sets of calibration curves: the first one 
obtained with proton showers, the second one from only iron events and, finally, taking both type of primaries together (called {\it mixed} here) as 
shown in Sec.~\ref{Sec:Results:Calibration}. The points and the error bars are the median and the confidence levels at $68$ and $95\%$ respectively.}
\label{Fig:EnergyBias_vs_Cp}
\end{figure}

We have determined the energy error as a function of the composition of the reconstruction sample. Since the energy error and resolution are almost 
energy independent for every primary (Fig.~\ref{Fig:ErecBias}), we have mixed all the events in this analysis regardless of their energy. 
We have used $100$ samples with $100$ events each. Proton and Iron primaries are randomly selected such the proton fraction varies from 
$0$ to $1$ in $0.1$ steps. The energy error as a function of the proton fraction is shown in Fig.~\ref{Fig:EnergyBias_vs_Cp}.  
The errors are also shown for the case that the calibration curves would have been obtained for each primary separately (we call them {\it proton}
and {\it iron} calibrations in the figure). As expected, the error is negligible if a sample with proton fraction of 0.5 is reconstructed with the calibration 
from Fig.~\ref{Fig:CalibCurves} since equal fraction of proton and iron events were used to get this calibration curves. Note also that if each primary 
is reconstructed with its own calibration the error is also very close to zero. In the unrealistic scenario that an extremely pure composition sample 
were reconstructed with the calibration obtained from a mixed composition flux, the absolute value of the incurred error would be  $<10\%$.

\subsection{Different array sizes and geometries}
\label{Sec:Results:Arrays}
The robustness of the method has been tested by varying the geometry of the array, i.e. the shape of the unitary cell and the distance between 
detectors. 

\begin{figure}
\centering
\includegraphics[width=8.1cm]{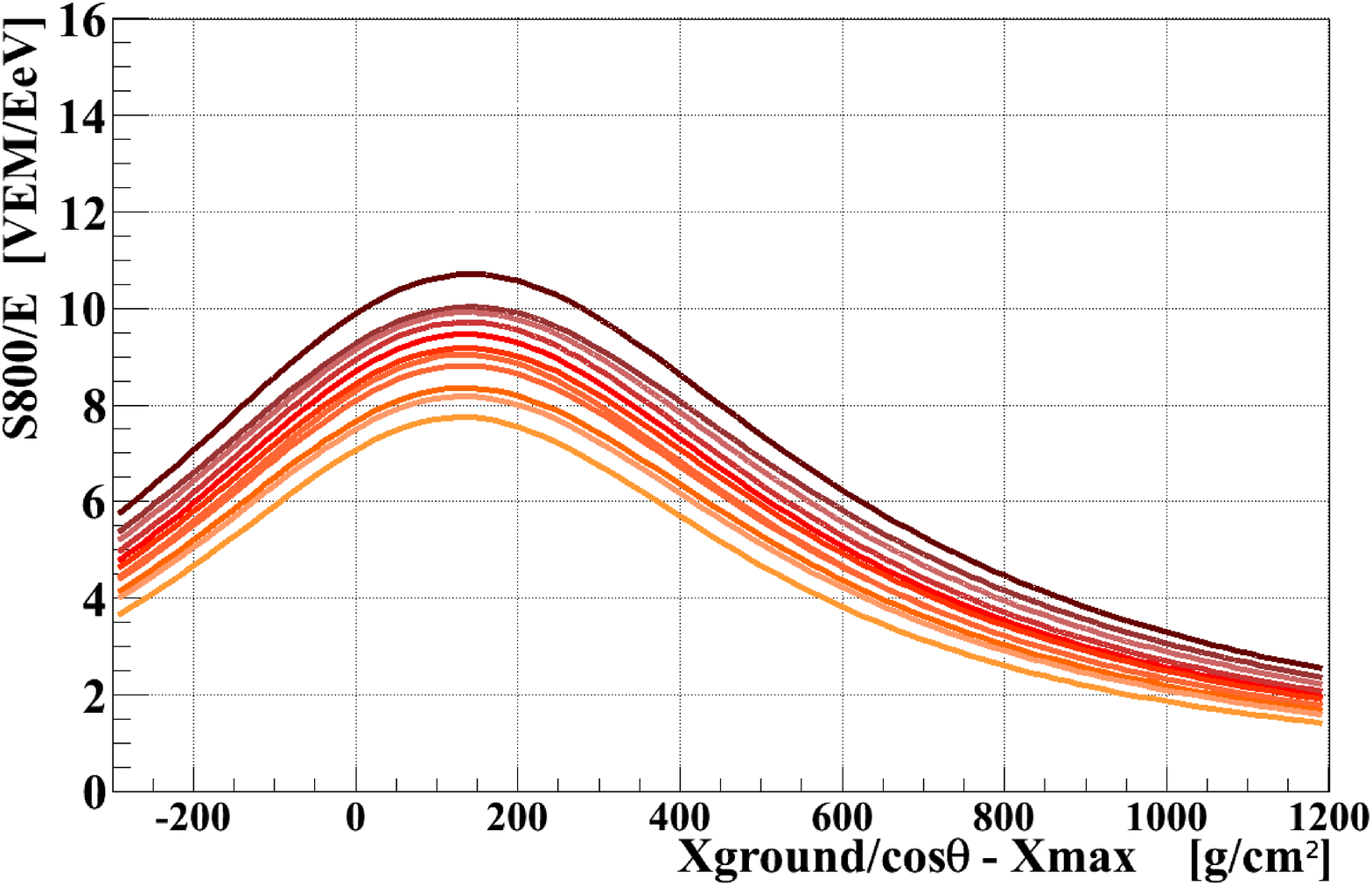}
\includegraphics[width=8.1cm]{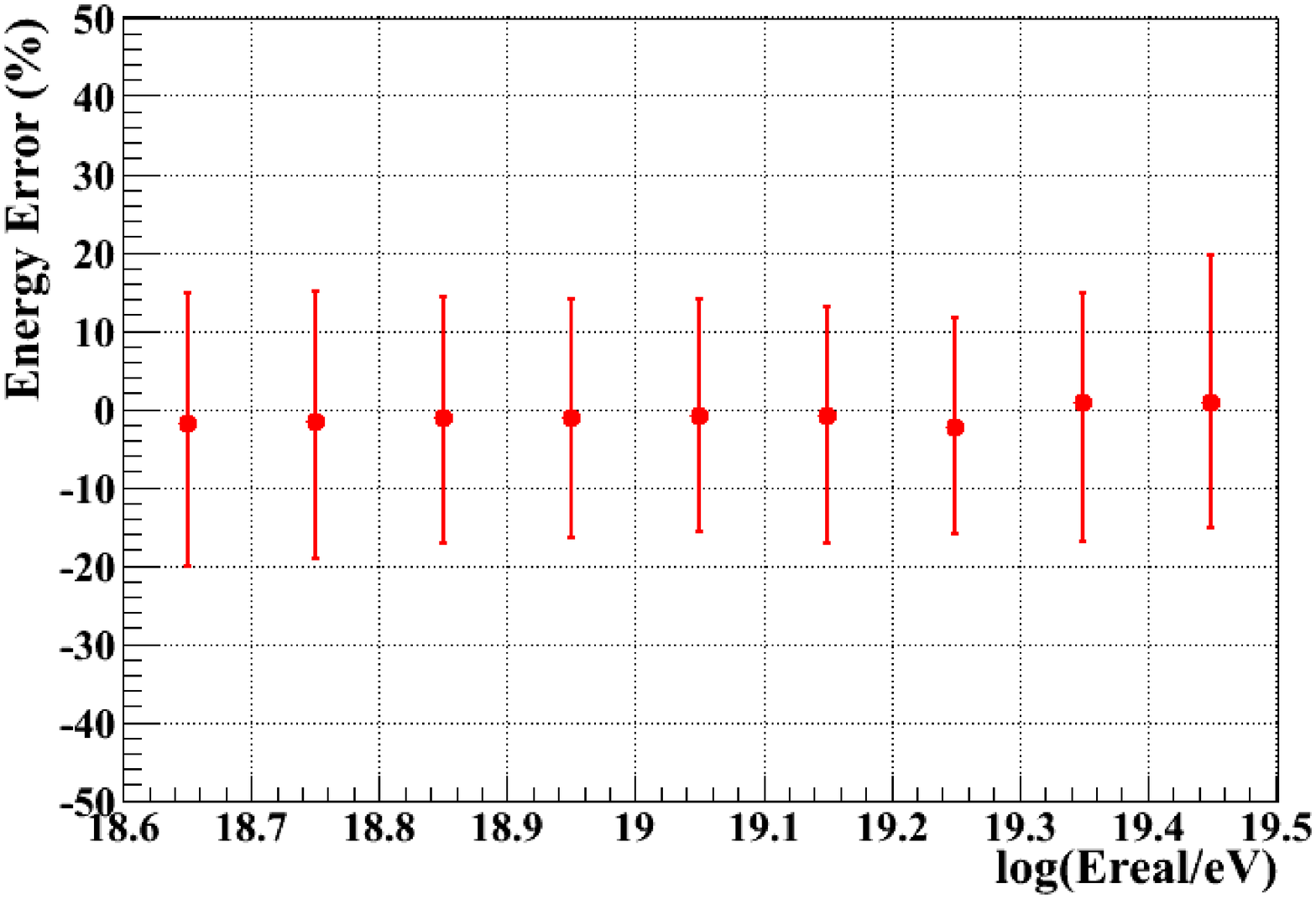}
\caption{Calibration curves (top) and error in the reconstructed energy (bottom) for the Telescope Array geometry. The points and the error bars 
are the median and the confidence levels at 68$\%$.}
\label{Fig:TA}
\end{figure}

First, an array with detectors arranged in a squared layout, with a separation of 1200 m is analyzed. Such a geometry is characteristic, for example, 
of the Telescope Array (TA) observatory. The applicability of the method to a scintillator array as TA is not analyzed here since it would require a 
different study as a consequence of the different response of scintillators to the muonic part of the shower compared to Cherenkov tanks. As mentioned, 
this Section focuses on its application to Cherenkov tank arrays with different geometries. The calibration curves and the error in the 
reconstructed energy are shown in Fig.~\ref{Fig:TA} for a mixed sample. Note that $S(800)$ has been selected as the energy estimator \cite{TASDSpectrum:13}. 
The error is energy independent and smaller than 1$\%$, while the resolution is $\sim$16$\%$.

\begin{figure}
\centering
\includegraphics[width=8.1cm]{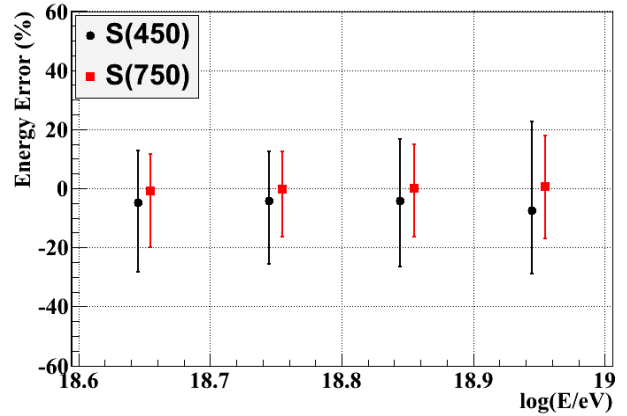}
\caption{Error in the reconstructed energy in case of a triangular array whose distance between detectors is 750 m. Two different energy estimators
are used, $S(450)$ and $S(750)$. Points and error bars are the median and 68$\%$ confidence levels. See text for details.}
\label{Fig:Infill}
\end{figure}

Second, a smaller triangular array is considered. A geometry representative of the Auger Infill is selected, i.e.~a triangular grid with 750 m spacing 
between detectors. The Infill is essentially designed to explore the region below the ankle and to reach full efficiency above $\sim$10$^{17.6}$ eV 
\cite{AugerICRCAmiga:13}. The fraction of saturated events is very large above 10$^{19}$ eV, so we only analyzed here the interval from 10$^{18.6}$ to 
10$^{19}$ eV. In the energy range of the experiment, the average optimum distance is $r_0 = 450$ m, so $S(450)$ is selected as the energy estimator 
\cite{AugerICRCAmiga:13}. Then, the error in the reconstructed energy is around -5$\%$ as shown in Fig.~\ref{Fig:Infill}. The error comes from the 
lack of events close to the maximum of the calibration curves which, as mentioned previously, is a key point in order to get a reliable fit and 
calibration. Since the optimum distance increases  with primary energy \cite{Ros:09}, the average r$_0$ is higher for the energy range analyzed here. 
Thus, the error could be minimized if a larger $r_0$ were selected. For example, using $S(750)$ the error would be negligible (Fig.~\ref{Fig:Infill}).

\section{Conclusions}

An iterative method, previously developed to infer the primary energy of photon-induced showers in pure surface arrays, has been modified to be 
applicable to hadron-initiated showers and tested assuming a hybrid experiment.  The method is based on a set of calibration curves that relate the 
surface energy estimator, $S(r_0)$, and the depth of maximum development of the shower, $X_{max}$, thanks to hybrid events. In pure surface arrays, 
a similar procedure could be performed but the calibration should rely on Monte Carlo (MC) parametrizations. However, it is important to note that MC 
parametrizations could be affected by the fact that the simulations do not reproduce properly the available experimental data 
\cite{AugerICRCMuonsSims:13}, a major advantage of the hybrid experiments.

The original method is based on the well-known universality of the electromagnetic component of the showers and the small number of muons produced
in the development of photon cascades. However, the significant fraction of the muon component for hadron-initiated showers, breaks the universality and makes 
necessary to implement a full set of calibration curves depending on the primary energy.

Our own simulation program of the detector response has been used. Shower to shower fluctuations and reconstruction uncertainties (core 
position, LDF fit and signal fluctuations) have been implemented. Primary energy and zenith angles go from 10$^{18.5}$ to 10$^{19.6}$ eV  
and from 0 to 60 degrees respectively. Several array geometries varying the shape of the unitary cell (triangular and square) and the distance 
between detectors (1500, 1200 and 750 m) have been studied. Considering a mixed sample of proton and iron primaries, the energy is determined 
with a negligible error and resolution around 16$\%$ in the full energy range analyzed.

Obviously, the same composition is expected for hybrid (used for the calibration) and pure surface events, so the energy of the latter could
be determined with error close to zero. However, in the extreme scenario that the reconstructed events present a pure composition, the energy 
error could achieve $\sim$10$\%$. This could be considered as the maximum uncertainty of the method due to the unknown composition of the 
primary flux. 

The energy error distributions are nearly Gaussian, an important point to get a more reliable reconstruction of the shape and position of rapidly 
varying spectral features since they are easier to manage when applying deconvolution techniques in the spectrum determination.

\section{Acknowledgments}

All the authors have greatly benefited from their participation in the Pierre Auger Collaboration and its profitable scientific 
atmosphere. Extensive numerical simulations were made possible by the use of the UNAM  super-cluster \emph{Kanbalam}, computational resources at
ICN-UNAM and the \emph{UAH-SPAS} cluster at the Universidad de Alcal\'a. We want to thank J. A. Morales de los R\'ios for the maintenance of the 
\emph{UAH-SPAS} cluster. LdP and MDRF also thank to ITeDA-CNEA for its hospitality during their stay. ADS is member of the Carrera del Investigador 
Cient\'ifico of CONICET, Argentina.

We also thank the support of the MICINN Consolider-Ingenio 2010 Programme under grant MULTIDARK CSD2009-00064, ASTROMADRID S2009/ESP-1496, 
and EPLANET FP7-PEOPLE-2009-IRSES. This work is partially supported by grant PAPIIT 107413, several grants from CONACyT either directly or through 
the Network of High Energy Physics (Red FAE), CONICET PIP 114-201101-00360 and ANPCyT PICT-2011-2223, Argentina.

\end{document}